\documentclass[a4paper]{jpconf-kb}
\usepackage{amsmath}
\usepackage{iopams}
\usepackage{natbib}
\usepackage{graphicx}

\usepackage{multind}

\begin{document}

\title{Topological constraints in magnetic field relaxation}

\author{S.\ Candelaresi$^{12}$}

\address{$^1$ NORDITA, KTH Royal Institute of Technology and Stockholm University,
\phantom{$^1$ }Roslagstullsbacken 23, SE-10691 Stockholm, Sweden}
\address{$^2$ Department of Astronomy, AlbaNova University Center,
\phantom{$^2$ }Stockholm University, SE-10691 Stockholm, Sweden}

\ead{iomsn@physto.se, iomsn1@gmail.com}


\begin{abstract}

Stability and reconnection
of magnetic fields play a fundamental role in
natural and man-made plasma.
In these applications the field's topology determines the stability
of the magnetic field.
Here I will describe the importance of one topology quantifier,
the magnetic helicity,
which impedes any free decay of the
magnetic energy.
Further constraints come from the
fixed point index
which hinders the field to relax into the Taylor state.

\end{abstract}

\section{Introduction}

Examples of topologically intriguing magnetic fields can be found both
in nature and in laboratory plasma.
During solar eruptions, ejections of hot plasma often appear in the shape
of pig tail like loops or sigmoidal structures
\citep{Canfield1999}.
These shapes are the product of strong magnetic fields which force the
charged particles to move along the field lines, which
themselves have a loop like shape.
In laboratory plasma topologically non-trivial magnetic fields are used to
shield the walls of the apparatus from the hot plasma.
In order to improve stability, the field is often twisted
\citep{Taylor1982} and can form even more sophisticated helical shapes,
such as those in the Large Helical Device in Toki, Japan \citep{Motojima2006LHD}.

\begin{figure}[!ht]
\begin{minipage}{0.3\linewidth}\begin{center}
\includegraphics[width=0.9\linewidth]{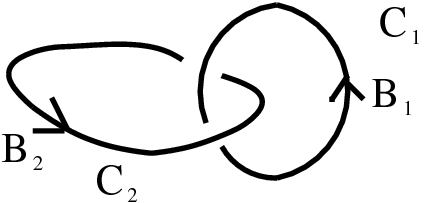}
\caption{\label{fig: two_rings}
Hopf link consisting of magnetic flux tubes.}
\end{center}\end{minipage} 
\begin{minipage}{0.3\linewidth}\begin{center}
\includegraphics[width=0.9\linewidth]{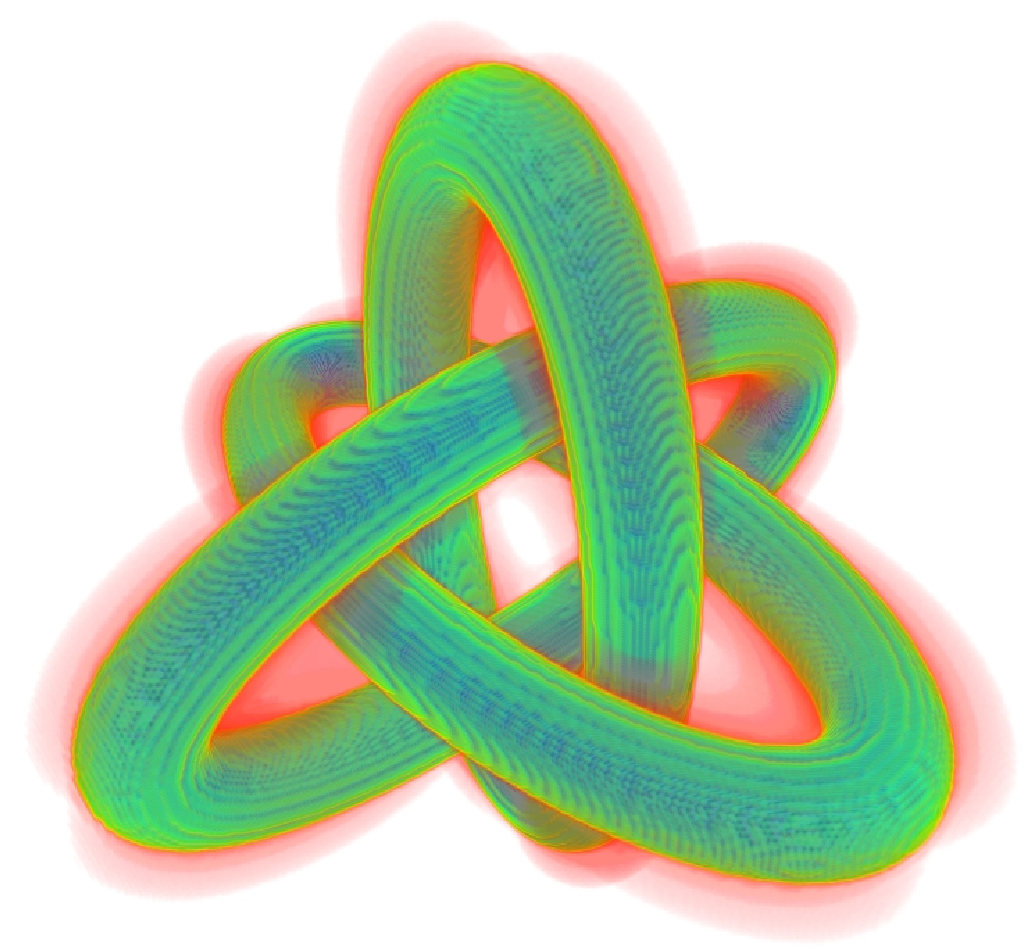}
\caption{\label{fig: Borromean}
Volume rendering of the initial magnetic energy for
the Borromean rings configuration.}
\end{center}\end{minipage}\hspace{2pc}%
\begin{minipage}{0.3\linewidth}\begin{center}
\includegraphics[width=0.9\linewidth]{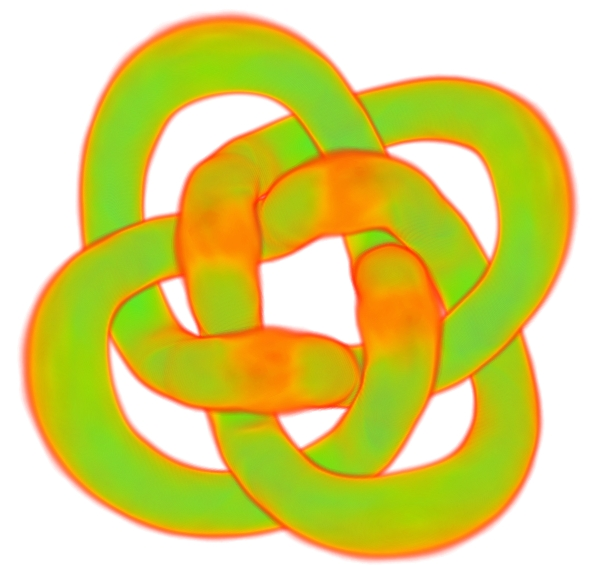}
\caption{\label{fig: iucaa}
Volume rendering of the initial magnetic energy for
the $8_{18}$ knot.}
\end{center}\end{minipage}\hspace{2pc}%
\end{figure}

During their evolution, astrophysical magnetic fields can
break up, reconnect or simply diffuse, which does not only alter
their geometry but in some cases also their topology.
Two magnetic vector fields are topologically different if one cannot be
transformed into the other without magnetic reconnection, i.e.\ without
breaking field lines and connecting them in a different manner.
Examples are magnetic flux tubes of the shape of
knots and
links
(Fig.~\ref{fig: two_rings}--Fig.~\ref{fig: iucaa}).

\section{Magnetic Helicity}

Magnetic helicity
$H=\int\boldsymbol{A}\cdot\boldsymbol{B}\ {\rm d}V$,
where $\boldsymbol{B} = \boldsymbol{\nabla}\times\boldsymbol{A}$ is the
magnetic field expressed as its vector potential $\boldsymbol{A}$,
is the most widely used measure
for qualifying and quantifying the field's topology.
It measures the number of mutual linkage of two magnetic field lines
\citep{MoffattKnottedness1969}.
This popularity arises from both from its universal applicability, i.e.\ it is useful
for delimited and diffusive fields, and from its conservation in ideal MHD.

In astrophysics and plasma physics one is often interested in the stability
of the field.
Magnetic helicity is approximately conserved in such systems,
suggesting that helical systems are more stable.
\cite{ArnoldHopf1974} captured this characteristic in the
realizability condition,
which expresses a lower limit for the magnetic energy in presence
of magnetic helicity
\begin{equation}
E(k) \ge k|H(k)|/(2\mu_{0}),
\label{eq: realizability}
\end{equation}
with the spectral magnetic energy $E(k)$, magnetic helicity $H(k)$,
the wave number $k$ and the vacuum permeability $\mu_{0}$.
It should be noted that, although magnetic helicity is generally gauge
dependent, the realizability condition is not, which preserves its physical
relevance.

Helical field configurations like the Hopf link (see Fig.~\ref{fig: two_rings})
consist of field lines which
are mutually interlocked or self interlinked.
Such linkage obviously obstructs the evolution of the field as long as magnetic
reconnection is not very strong.
This resistance against free evolution is captured in the realizability condition
\eqref{eq: realizability}.
There are, however, field configurations of interlinked field lines which are
not helical, like the Borromean rings
(Fig.~\ref{fig: Borromean})
and the $8_{18}$ knot (Fig.~\ref{fig: iucaa}).
Are such magnetic links and knots restricted in their evolution because of
their linkage, or are they free, because the realizability condition does not
apply?

\begin{figure}[!ht]
\begin{center}
\includegraphics[width=0.1\linewidth]{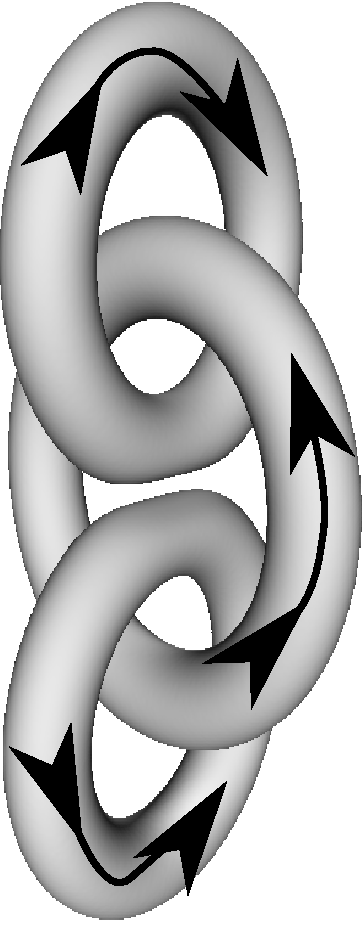}
\hspace{.18\linewidth}
\includegraphics[width=0.1\linewidth]{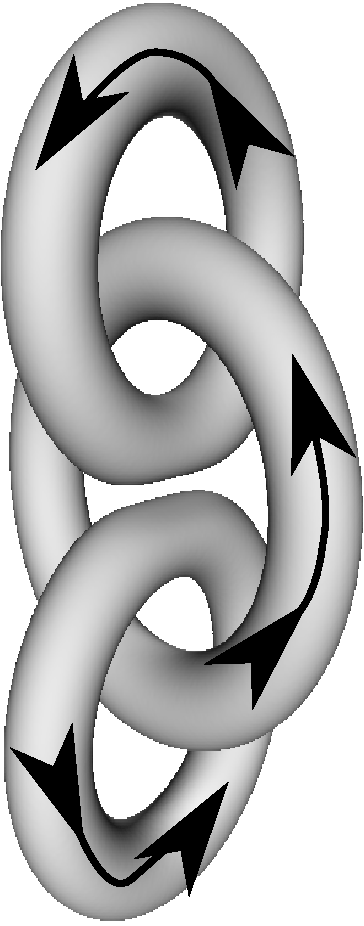}
\hspace{.18\linewidth}
\includegraphics[width=0.12\linewidth]{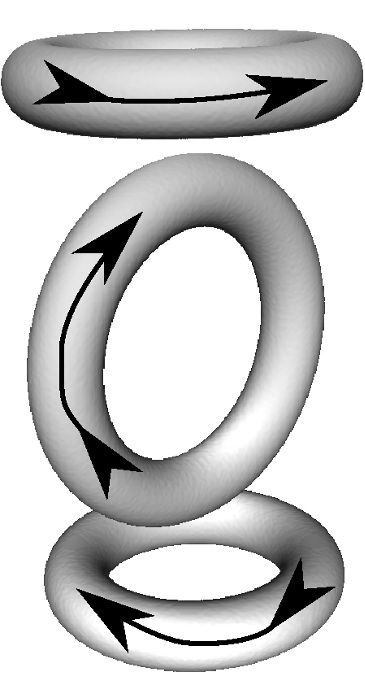}
\caption{\label{fig: three rings}
Iso surfaces for the initial magnetic field for the triple ring
configurations.
The left configuration is helical, while the other two are non-helical.
Taken from \citep{fluxRings10}.
}
\end{center}
\end{figure}

\begin{figure}[!ht]
\begin{minipage}{0.49\linewidth}
\begin{center}
\includegraphics[width=\linewidth]{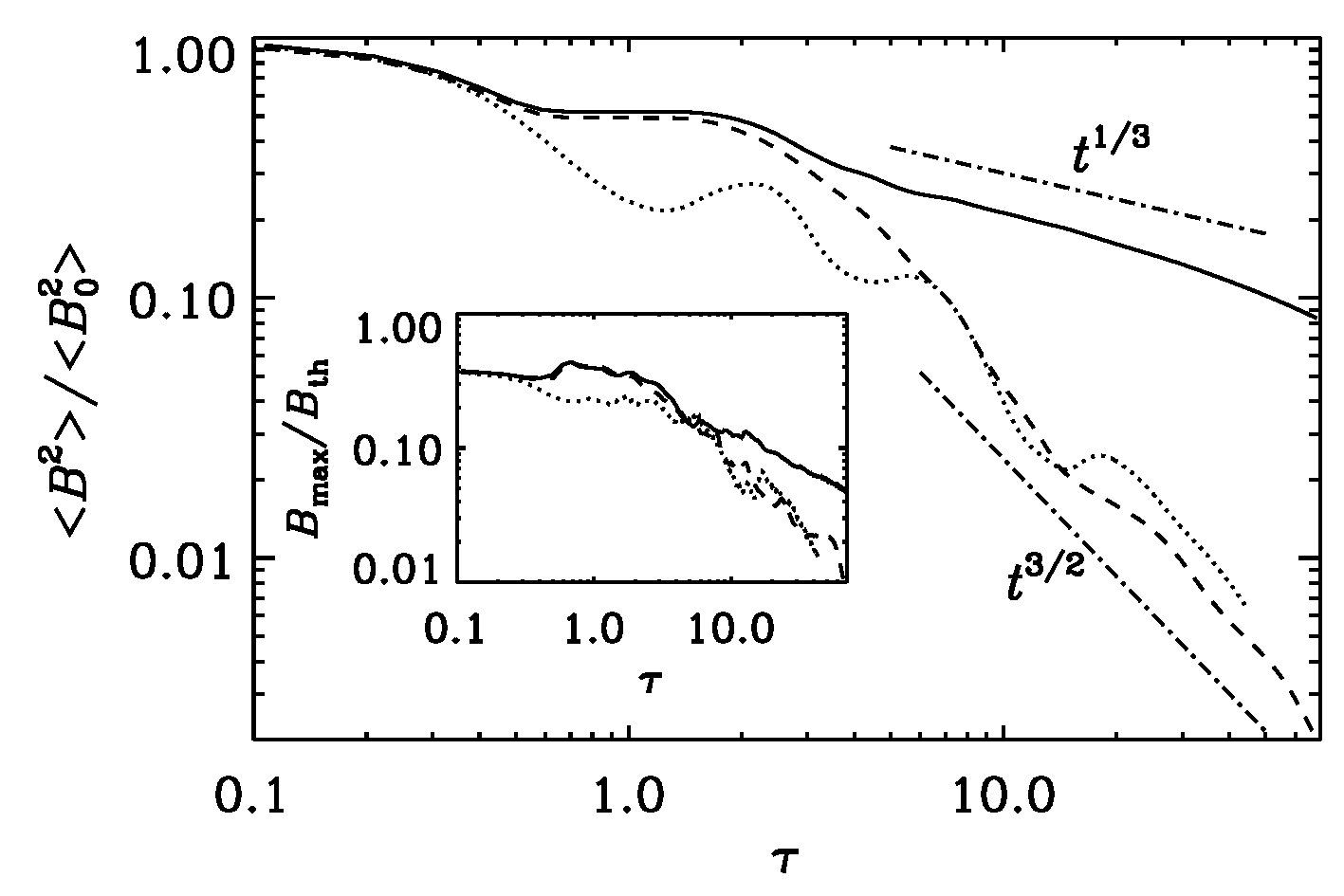}
\caption{\label{fig: energies rings}
Normalized magnetic energy in time for the helical triple rings (solid line),
the non-helical interlocked rings (dashed line) and un-linked rings
(dotted line).
The linking of field lines is not able to slow down the decay of
magnetic energy.
On the other hand, the presence of magnetic helicity poses a
constraint in the field's relaxation.
Taken from \citep{fluxRings10}.
}
\end{center}
\end{minipage}
\hspace{.01\linewidth}
\begin{minipage}{0.49\linewidth}
\begin{center}
\includegraphics[width=\linewidth]{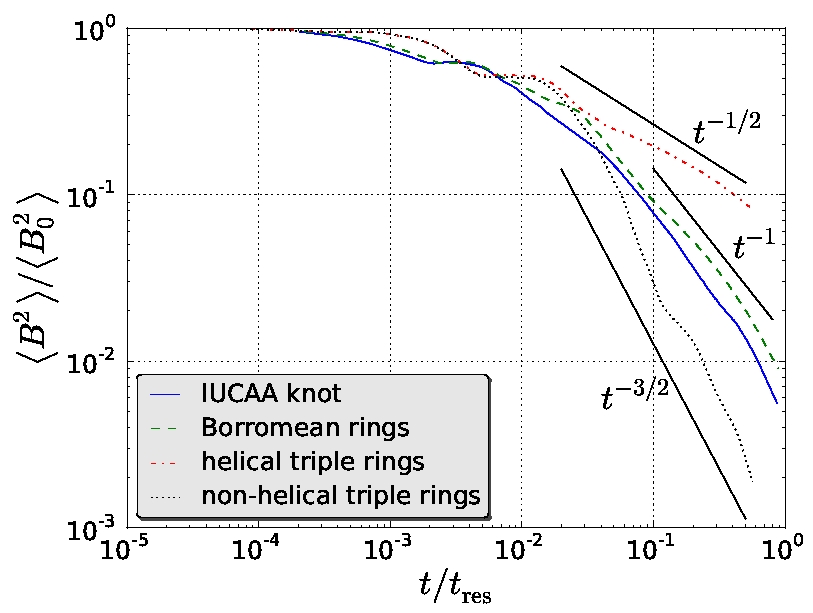}
\caption{\label{fig: energies compare}
Comparison of the magnetic energy evolution for the interlocked triple
rings, the Borromean rings and the IUCAA knot ($8_{18}$ knot).
The non-helical Borrmean rings and the IUCAA knot show an intermediate
decay behavior, of which the first one can be explained by the occurrence
of separated helical structures.
Taken from \citep{knotsDecay11}.
}
\end{center}
\end{minipage}
\end{figure}

In a resistive and viscous plasma it is easy to simulate the decay of magnetic
knots and links \citep{fluxRings10}.
Three initial magnetic field configurations are considered, each comprising three
flux rings (Fig.~\ref{fig: three rings}).
The first and second configurations are linked, the first of which is helical.
The third un-linked setup servers for comparison.
Initially all three setups show a very similar decay in magnetic energy
(Fig.~\ref{fig: energies rings}).
After about 10 Alfvenic times, however, they decay distinctively.
For the helical setup we observe a rather slow decay in magnetic energy, which
for late times, behaves like a power law of $t^{-1/3}$.
For the two non-helical configurations this power law is steeper and of the order
of $t^{-3/2}$.
Unlike it might be assumed, the actual linking of the rings does not affect
the dynamics considerably.
Magnetic helicity,
on the other hand, restricts the relaxation such that
magnetic energy decays only slowly.
The two non-helical cases decay equally fast and the significance of the
realizability condition
is confirmed.

\section{Beyond magnetic helicity}

\begin{figure}[!ht]
\begin{center}
\includegraphics[width=0.5\linewidth]{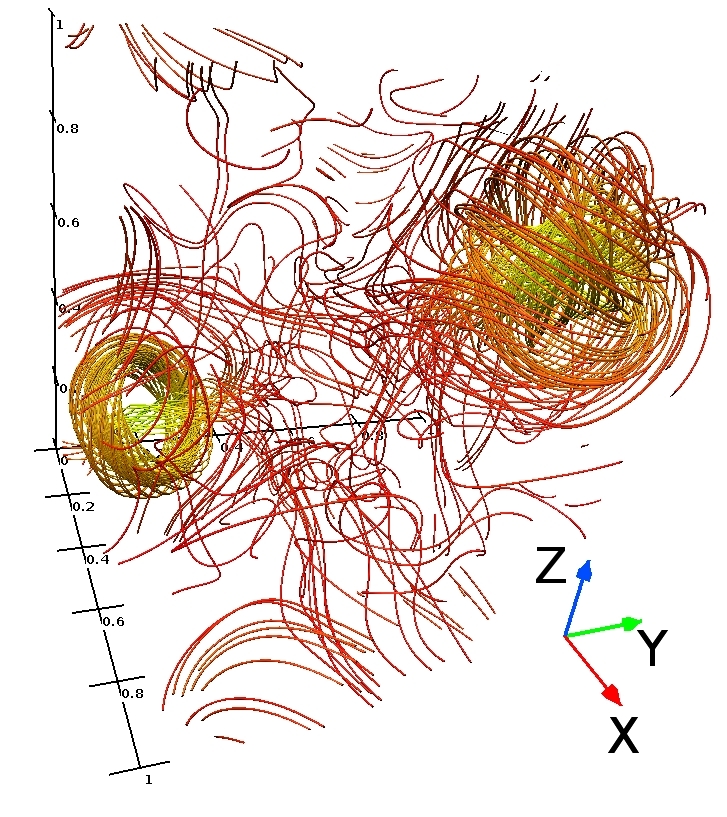}
\caption{\label{fig: Borromean field lines}
Magnetic streamlines of the Borromean rings configuration after a decay over
78 Alfvenic times. The colors represent the strength of the magnetic field.
Two helical structures appear which are spatially separated.
Their helicity content impedes any fast decay of magnetic energy.
Taken from \citep{knotsDecay11}.
}
\end{center}
\end{figure}

Other than the magnetic helicity there exist third and forth order topological
invariants which are non-zero for the
Borromean rings \citep{ruzmaikin:331}.
Whether such invariants affect the field's evolution is tested in comparison with
the triple rings (Fig.~\ref{fig: energies compare}) \citep{knotsDecay11}.
Reconnection causes the Borromean rings to reshape into two twisted tori
(Fig.~\ref{fig: Borromean field lines}).
Those tori are helical, and so the realizability condition has an effect
locally, which can be observed in the slowed down decay of the magnetic energy
(Fig.~\ref{fig: energies compare}).
Any higher order topological invariant is not needed to explain the intermediate
decay, but cannot be excluded either.

Perhaps the most successful application of topological invariants other than
the magnetic helicity is the fixed point
index \citep{Brown1971}.
\cite{Yeates_Topology_2010} showed
that its conservation imposes an additional restriction
in the field's evolution such that the relaxed state reaches higher energies than
proposed by \cite{Taylor-1974-PrlE}.
Any application of the fixed point index on particularly non-helical knots
and links in the form of braids would give considerable new insights in
their relaxation dynamics.

\section{Conclusions}

We have shown that magnetic helicity, rather than the topology, is the most
important parameter in determining the evolution of magnetic links and knots.
Nevertheless, a non-helical initial field can spontaneously
transform into separate helical fields which are restricted in their evolution
(see Borromean rings).
Combining the fixed point index with various topologically non-trivial knots
and links would enhance our understanding of topological constraints and
magnetic reconnection.



%

%
\ack

We thank Dr.\ Anthony van Eysden for suggestions in improving the text's
quality.
We also thank the Swedish National Allocations Committee for
providing computing resources at the National Supercomputer
Centre in Link¨oping and the Center for Parallel Computers
at the Royal Institute of Technology in Sweden.
This work
was supported in part by the Swedish Research Council, Grant
621-2007-4064, the European Research Council under the AstroDyn
Research Project 227952, and the National Science
Foundation under Grant No. NSF PHY05-51164.

\bibliographystyle{jfm2}

\bibliography{references}
\end{document}